\documentclass[preprintnumbers,two column, showpacs,amsmath]{revtex4}
\usepackage{graphicx}
\usepackage{dcolumn}
\usepackage{bm}

\begin{document}

\title{Jaynes-Cummings model:Counter rotating effect on the vacuum Rabi splitting
and atom-cavity dynamics}
\author{Yu-Yu Zhang$^{1,2}$, Qing-Hu Chen$^{3,4, *}$, and Shi-Yao Zhu$^{1,5, \dagger
}$}
\date{\today}

\address{
$^{1}$Beijing Computational Science Research Center, Beijing
100084, P. R. China  \\
$^{2}$Center for Modern Physics, Chongqing University£¬ Chongqing
400044, P. R.  China\\
$^{3}$Center for Statistical and Theoretical Condensed Matter
Physics, Zhejiang Normal University, Jinhua 321004, P. R. China  \\
$^{4}$Department of Physics, Zhejiang University, Hangzhou 310027,
P. R. China \\
$^{5}$Department of Physics, Hong Kong Baptist University, Hong
Kong, P. R. China
 }

\begin{abstract}
The effect of the counter-rotating terms in the Jaynes-Cummings
model is investigated with an extended coherent-state approach. The
counter-rotating terms greatly modify the vacuum Rabi splitting. Two
peaks with different heights in the weak coupling regime and more
than two peaks in the intermediate coupling regime are predicted.
The collapses and revivals in the evolution of the atomic population
inversion disappear in the intermediate coupling regime, but
reappear in the strong coupling regime. This reappearance is similar
to that under the rotating-wave approximation, attributed to the
summation of periodic cosine functions of the evolution.
\end{abstract}


\maketitle

\address{
$^{1}$Beijing Computational Science Research Center, Beijing
100084, P. R. China  \\
$^{2}$Center for Modern Physics, Chongqing University£¬ Chongqing
400044, P. R.  China\\
$^{3}$Center for Statistical and Theoretical Condensed Matter
Physics, Zhejiang Normal University, Jinhua 321004, P. R. China  \\
$^{4}$Department of Physics, Zhejiang University, Hangzhou 310027,
P. R. China \\
$^{5}$Department of Physics, Hong Kong Baptist University, Hong
Kong, P. R. China
 }

\address{
$^{1}$Beijing Computational Science Research Center, Beijing
100084, P. R. China  \\
$^{2}$Center for Modern Physics, Chongqing University£¬ Congqing
400044, P. R.  China\\
$^{3}$Center for Statistical and Theoretical Condensed Matter
Physics, Zhejiang Normal University, Jinhua 321004, P. R. China  \\
$^{4}$Department of Physics, Zhejiang University, Hangzhou 310027,
P. R. China \\
$^{5}$Department of Physics, Hong Kong Baptist University, Hong
Kong, P. R. China
 }

The matter-light interaction is a fundamental one in optical
physics.  The simplest model is the Jaynes-Cummings (JC)
model\cite{jaynes}, which describes the interaction between a
two-level atom and a single mode of the quantized electromagnetic
field. In the semiclassical description, the evolution of the atomic
population inversion shows a regular oscillation behavior. In the
quantum mechanics description with the rotating-wave
approximation(RWA), JC model is exactly solveable\cite{Scully}. By
the exact solution, the revival and collapse of the atomic
population inversion has been predicted by Eberly et
al\cite{orszag,eberly,eberly1}, and later confirmed
experimentally\cite{rempe}. The other important quantum mechanical
feature is the vacuum Rabi splitting~\cite{Thompson,kimble}, which
is equal to 2g (g the atom-cavity coupling strength) in the JC model
with the RWA.

The RWA is valid in the weak coupling regimes with small detuning in
the study of the JC model, because the  contribution of the
counter-rotating terms (CRTs) to the evolution of the system is
quite small. In the coupling is not weak, the CRTs, which allow the
simultaneous creation and annihilation of an excitation quantum in
both atom and cavity mode, must be taken into account. The CRTs
would modify the above two interesting quantum phenomena, i.e. the
revivals and collapses of the atomic population inversion and the
vacuum Rabi splitting. A perturbation by path-integral method has
been applied to study the contribution of counter-rotating terms in
quantum Rabi oscillation in intermediate-coupling
regime\cite{zubairy}. The effect of the CRTs in strong-coupling
regime  with large detuning was discussed in Ref. ~\cite{casanova}
by using a perturbation method .

Recently, JC model in the intermediate coupling regime has been
realized in circuit quantum electrodynamics (QED) where the
superconducting qubits play the role of artificial
atoms~\cite{Abdumalikov,Fink,Niemczyk,prorn,Fedorov}. By enhancing
the inductive coupling, the atom-cavity coupling strength reaches a
considerable fraction (maximum is around  $12\%$ ) of the cavity
transition frequency\cite{Niemczyk}. Theoretically, some approaches
are proposed toward the accurate solution to the JC model without
the RWA\cite{Irish,Zucco,Ashhab,Hwang,chen,chen1}. Among them, the
effective approaches within extended bosonic coherent states have
been developed \cite{chen,chen1}.

In this Letter, we study the JC model without the RWA beyond the
weak coupling regime by the numerically exact solutions\cite{chen}.
We consider the resonant case which is important in the interaction
of light and the matter. The effect of the CRTs on the dynamic
evolution   of the atomic population inversion and the vacuum Rabi
splitting is investigated.

The Hamiltonian of a two-level atom with transition frequency $
\omega_{eg}$ interacting with a single-mode quantized cavity of
frequency $ \omega$  is
\begin{equation}  \label{ham}
H=\frac{\omega_{eg}}{2}\sigma_{z}+\omega
a^{\dagger}a+g(a^{\dagger}+a) \sigma_{x},
\end{equation}
where $g$ is coupling strength, $\sigma_x$ and $\sigma_z$ are Pauli
spin-$1/2$ operators, $a^{\dagger}$ and $a$ are the creation and
annihilation operators for the quantized field.

By using the extended coherent state
technique\cite{chen, chen1}, we may write a transformed Hamiltonian
with a rotation around an $y$ axis by an angle $\frac \pi 2$
\begin{equation}  \label{ham1}
H_U=-\frac{\omega_{eg}}{2}\sigma_{x}+\omega
a^{\dagger}a+g(a^{\dagger}+a) \sigma_{z}.
\end{equation}
Then we  introduce two new boson operators $A=a+g/\omega$ and
$B=a-g/\omega$ to eliminate the linear term. The Fock states with
occupation number $m$ in $A$ and $B$ are
\begin{equation}\label{coherent1}
|m\rangle_{A(B)}=(a^{\dagger}\pm
g/\omega)^{m}e^{\mp\frac{g}{\omega}
a^{\dagger}-g^2/2\omega^2}|0\rangle,
\end{equation}
The wavefunction of the transformed Hamiltonian can be expanded as
\begin{eqnarray}  \label{wavefunction}
|\varphi \rangle=\sum_{m=0}^{N_{tr}}c_{m,1}|m\rangle_A|1\rangle
+c_{m,2}|m\rangle_B|2\rangle,
\end{eqnarray}
where $|1\rangle=\frac{1}{\sqrt{2}}(|e\rangle+|g\rangle)$ and
$|2\rangle=\frac{1}{\sqrt{2}}(-|e\rangle+|g\rangle)$ with
$|e\rangle$ and $|g\rangle$  the upper and low atomic state, $
N_{tr}$ is the truncated number. The wavefunction in original photon
Fock space can also be expressed as
\begin{eqnarray}  \label{fock state}
|\varphi \rangle&=&\frac{1}{\sqrt{2}}\sum_{n,m}\langle
n|(c_{m,1}|m\rangle_{A}-c_{m,2}|m\rangle_{B})|n,e\rangle\nonumber\\
&+&\langle
n|(c_{m,1}|m\rangle_{A}+c_{m,2}|m\rangle_{B})|n,g\rangle
\end{eqnarray}

\begin{figure}[tbp]
\includegraphics[width=9cm]{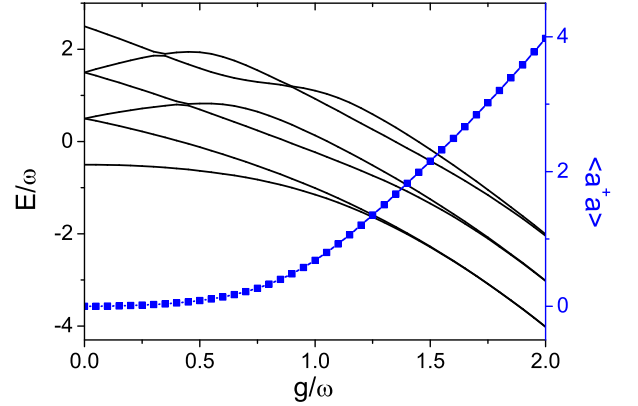}
\caption{The first six energy levels (black solid line) and the mean
photons number of the ground states $a^{+}a$ of JC model with
CRTS(blue squared line) for atom and cavity on resonance
$\omega=\omega_{eg}$. } \label{energy}
\end{figure}

Associated with JC Hamiltonian with CRTS is a conserved parity
$\prod=e^{-i\pi\hat{N}}$ with the excitation number
$\hat{N}=a^{\dagger}a+\frac{1}{2}\sigma_z+\frac{1}{2}$. $\prod$
possesses two eigenvalues $\pm1$, depending on whether the
excitation quanta number is even or odd. The coefficients therefore
satisfy $c_{m,2}=\pm (-1)^{m}c_{m,1}$ and the wavefunction with even
or odd parity can be written as
\begin{eqnarray}  \label{wavefunction}
|\varphi^{\pm} \rangle=\sum_{m=0}^{N_{tr}}c_{m,1}|m\rangle_A|1\rangle
\pm (-1)^{m}c_{m,1}|m\rangle_B|2\rangle,
\end{eqnarray}
The coefficients, $c_{m,1}$ and the corresponding energies, $E^{\pm}$, are determined by
\begin{eqnarray}  \label{equation1}
E^{\pm}\sum_{m=0}^{N_{tr}}c_{m,1}|m\rangle_A &=&\mp\frac{\omega_{eg}}{2}\sum_{m=0}^{N_{tr}}(-1)^{m}c_{m,1}|m\rangle_B|g\rangle \nonumber \\
&+&\omega\sum_{m=0}^{N_{tr}}(A^{\dagger}A-g^2/\omega^2)c_{m,1}|m\rangle_A.\nonumber \\
\end{eqnarray}
The solutions can be obtained by digonalizing the set of equations
with dimension $ M=N_{tr}+1$. The truncated number $N_{tr}$ is set
to $40$ in the present numerical calculations, yielding the
eigenenergies with relative errors less than $10^{-6}$. The energies
of the first six eigenstates and the average photon number in
the ground state versus $g/\omega$ are plotted in Fig.~\ref{energy}.
Here and in the following calculation, we take the resonant case
$\omega_{eg}=\omega$

The eigenfunction $|\varphi^\pm\rangle$ can be expressed in the
states $|n,e\rangle$ and $n,g\rangle$ by using the relation
$\langle n|m\rangle_A=(-1)^{n}D_{nm}$ and $\langle
n|m\rangle_B=(-1)^{m}D_{nm}$ with polynomials
$D_{nm}=\sum_{i=0}^{min[n,m]}(-1)^{i}\frac{\sqrt{n!m!}}{i!(m-i)!(n-i)!}(g/\omega)^{m+n}e^{-g^2/2\omega^2}$,
for the even parity
$$
|\varphi^{+}\rangle=\sum_{n=odd,m}-c_{m,1}D_{nm}|n,e\rangle
+\sum_{n=even}c_{m,1}D_{nm}|n,g\rangle
$$
and for the odd parity
$$
|\varphi^{-}\rangle=\sum_{n=even,m}c_{m,1}D_{nm}|n,e\rangle
-\sum_{n=odd}c_{m,1}D_{nm}|n,g\rangle.
$$

\begin{figure}[tbp]
\includegraphics[width=9cm]{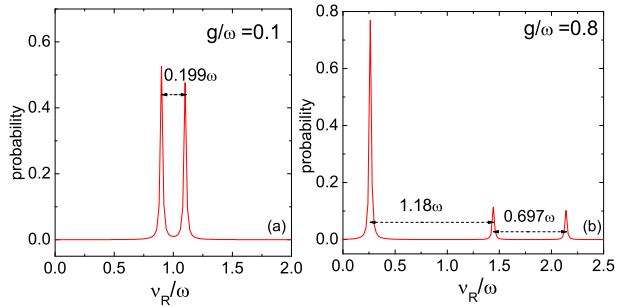}
\caption{(Color online) Emission spectrum $\nu_{R}/\omega$ the
initial excited state $V|gs\rangle$ with $\omega=\omega_{eg}$ for
(a) $g/\omega=0.1$, (b) $g/\omega=0.8$. } \label{frequency}
\end{figure}

Based on the above new states, Eq.(~\ref{coherent1}), we study
the influence of the CRTs on the Rabi splitting, particularly the
vacuum Rabi splitting. When the atom dressed by the cavity mode is
subjected to the interaction with the vacuum modes, we would have
spontaneous emission if we pump the dressed atom from its ground
to an excited state. Under the RWA, when the atom is excited by
the operator, $V=|e\rangle\langle g|+|g\rangle\langle e|$, from
the ground state $|g,0>$, the emission spectrum has two peaks with
equal height (the distance of the two peaks, $2g$, is the vacuum
Rabi splitting). When the CRTs are included, the photons in the
ground state are no longer a vacuum state, as shown in
Fig.~\ref{energy}. Here, we still use $V$ to excite the atom from
the ground state of the atom-cavity system, and regard the
splitting in the emission spectrum as the vacuum Rabi splitting.
The excited state, $V|gs\rangle$, can be expanded in terms of the
eigenstates with odd parity
\begin{eqnarray}
V|gs\rangle&=&(|e\rangle\langle g+|g\rangle\langle
e|)|\varphi^{+}_{0}\rangle\nonumber\\
&=&\sum_{k}|\varphi^{-}_{k}\rangle\langle\varphi^{-}_{k}[\sum_{n=odd,m}-c_{m,1}D_{nm}|n,g\rangle\nonumber\\
&+&\sum_{n=even}c_{m,1}D_{nm}|n,e\rangle].
\end{eqnarray}
From the excited state, the atom will decay to the dressed ground
state with an emission spectrum. The spectrum has Lorenzian peaks
due to the spontaneous transitions between high energy eigenstate
components in the excited state to the dressed ground state, and the
heights of the peaks is proportional to the square of the
probability of the corresponding eigenstates. The width of the peaks
is the same determined by the decay rate. The frequency
difference(s) will give us the vacuum Rabi splitting. In
Fig.~\ref{frequency}, we present the emission spectrum for different
coupling strength $g$ at resonance $\omega_{eg}$=$\omega$.
Three new features are obviously observed.

$\bullet$ At small $g/\omega=0.1$, there are two peaks, but their
heights are not the same.

$\bullet$ More than $2$ peaks can emerge in the emission spectrum,
as shown in Fig.~\ref{frequency} (b) for $g/\omega=0.8$ with $3$
peaks. The distance between peaks can be different. The first
splitting is $1.18\omega$
and the second splitting is $0.697\omega$.

$\bullet$ At the strong coupling, $g/\omega =2$, there is no
Rabi splitting, which is the result of almost equal energy
separation for all eigenstates. The energy difference of the ground
state and the first-excited state is $0.0004\omega$ obtained by
Eq.(~\ref{equation1}) numerically, and the transition between them
is smeared out for temperature $T>3\times10^{-15}\omega$.

Without the interaction between atom and the cavity mode, we have
energy degeneracy for $|e,n\rangle$ and $|g,n-1\rangle$ , and no
degeneracy for $|g,0\rangle$, which results in three peak structure
for $n\neq0$ and the vacuum Rabi splitting  (two peak structure) for
$n=0$. The degeneracy is broken by the interaction. For small
$g/\omega$ , the non-degeneracy is usually kept for all energies
under and without the RWA, and the eigenenergies without RWA are
close to that under the RWA. For $g/\omega>0.1$, the difference
between RWA and without RWA merges. For very large $g/\omega$, the
degeneracy reappears with different fashion: almost the same energy
for $|\varphi^{+}_{i}\rangle$ and $|\varphi^{-}_{i}\rangle$. The
vacuum Rabi splitting disappears. At the same time, the multi-peak
structure for $n\neq0$ also disappears.

Next, we study the effect of the CRTs on the dynamical evolution of the system.
Under the RWA we have the well-known collapse and revival. It has been shown
that the collapse and revival would become not clear when the
CRTs are included ~\cite{zubairy}. We investigate the effect of the CRTs by increasing
$g/\omega$ with the initial state, the atom in the ground state $|g\rangle$
and the field in a coherent state $ |\alpha\rangle$ with the mean photons
$\bar{n}=\langle\alpha|a^\dagger a|\alpha\rangle$.

The time evolution of the wavefunction is expressed as
\begin{eqnarray}  \label{dynamics wavefunction}
|\varphi (t)\rangle=e^{-iHt}|\varphi (0)\rangle
=U^{\dagger}\sum_{k}f^{\pm}_{ k}e^{-iE^{\pm}_{ k}t}|\varphi^{\pm}_{
k}\rangle,
\end{eqnarray}
where the coefficients are evaluated by $f^{\pm k}=\langle\varphi^{\pm}_{ k}|U|\varphi(0)%
\rangle$. By tracing over the field degrees of the freedom, the
evolution of the reduced density matrix for the atom  is given by
\begin{equation} \label{reduced matrix}
\hat{\rho}_{a}(t) = \sum_{k,l}^{M}f^{\pm}_{ k}f^{\pm}_{
l}e^{-i(E^{\pm}_{ k}-E^{\pm}_{
l})t}\mathtt{Tr}_{ph}(U^{\dagger}|\varphi^{\pm}_{ k}\rangle\langle
\varphi^{\pm}_{ l}|U)\nonumber \\.
\end{equation}
The atomic population inversion can be expressed as
\begin{equation}  \label{population}
P(t)=|\mathtt{Tr}((|e\rangle\langle
e|\hat{\rho_{a}}(t))|-|\mathtt{Tr}((|g\rangle\langle
g|\hat{\rho_{a}}(t)).
\end{equation}
with the probability $P_{j}=|\langle j|\hat{\rho}_a(t)|j\rangle|$
for the atom in the upper $j=1$ (down $j=2$) state at time $t$.
For convenience, we also write the population inversion in the
case of the RWA~\cite{eberly,eberly1,orszag}
\begin{equation}\label{PRWA}
P(t)\propto\sum_{n}\cos (2gt\sqrt{n+1}).
\end{equation}
Note that the collapse and revival evolution  is independent of $g$
in the dimensionless time scale of $\tau=2gt$.

\begin{figure}[tbp]
\includegraphics[width=8cm]{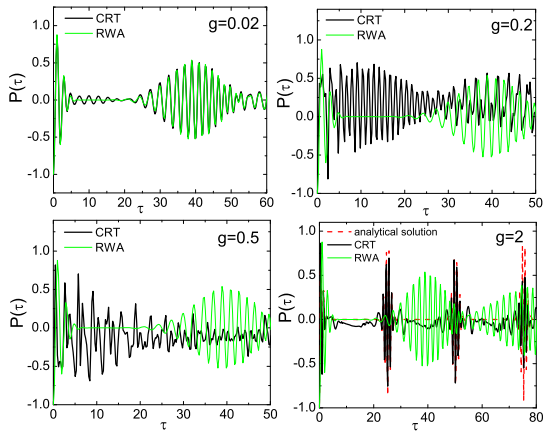}
\caption{ Population inversion evolution 7.  (solid curves) for the
atom-cavity system without RWA for different coupling strength
$g/\omega=0.02$ (a), 0.2 (b), 0.5 (c) and 2.0 (d). For comparison,
the corresponding RWA results (green curves), and the approximated
result from Eq. (\ref{analytical}) (red curves) are also plotted.
The initial mean photons $\bar{n}=10$ and
$\omega_{eg}=\omega$. } \label{comp}
\end{figure}

By using the accurate eigenvalues $\{E^{\pm}_{ k}\}$ and eigenstates
$\{\varphi ^{\pm}_{ k}\}$ and initial mean photon $\bar{n}=10$, we
calculate the population inversion evolution $P(\tau )$
without the RWA, see the black curves in Fig.~\ref{comp}
for different $g/\omega$. For comparison
the RWA results are also plotted, see the green curves. In
the weak coupling regime, say $g/\omega=0.02$, the collapse and
revivals can still be seen, with some oscillation especially
in the collapse period.
Due to the CRTs, the dependence of $2g\sqrt{n}$ in Eq. (11) is broken.
Therefore, as the increase of $g/\omega$ from $0.01$ to $1.0$, the
collapse and revivals will gradually disappear. In
Fig.~\ref{comp}(b) for $g/\omega=0.2$,
rapid oscillation arises in the collapse period
and the envelope in the revival period is smeared, which is
consistent with perturbation results
~\cite{zubairy}. As $g/\omega$ increases, the effect of the
CRTs becomes dominated. At $g/\omega =0.5$, the
collapse and revival completely disappear due to the CRTs, see
Fig.~\ref{comp}(c). The situation becomes quite different if
$g/\omega$ further increases. For large $g/\omega=2$, as shown in
Fig. 3(d) that the regular collapses and revivals reappear, which
will be explained below.

We now present the perturbation theory in the strong coupling
regime. Note that, without the first term
$H_{1}=-\frac{\omega_{eg}}{2}\sigma_{x}$,  the Hamiltonian
(~\ref{ham1}) can be diagonalized  exactly in terms of the new
operators $A$ and $B$. The $k$th eigenstate and eigenvalue read
\begin{equation}\label{strong-state}
|k^{\pm}>=\frac 1{\sqrt{2}}\left(
\begin{array}{l}
|k_A> \\
\pm (-1)^k|k_B>
\end{array}
\right), E^{\pm}_{k}=\omega k-g^2/\omega.
\end{equation}
Regarding $H_{1}$ as a perturbation term, we then obtain the
second-order perturbative results for  energy levels   as
\begin{equation}  \label{perturbed energy}
E^{\pm}_{k}=\omega k-g^2/\omega \mp\frac{\omega_{eg}}{4} L_{kk}^{}+\frac 14(%
\frac{\omega_{eg}}{2})^2\sum_{l\neq k}\frac{|L _{lk}^{}|^2}{\omega (l-k)},
\end{equation}
with polynomials $L_{lk}$ dressed by overlap of two coherent states as
$L_{kl}=\langle k_A|l_B\rangle(-1)^{l}+(-1)^{k}\langle k_B|l_A\rangle$.

By using the above unperturbated  eigenstate and  eigenvalues, the
probability of the state with the atom in state $|i\rangle$
($i=e,g$) and $n$ photons in the cavity mode is given by
\begin{eqnarray}
P_{n,1(2)}=\frac 14\sum_m|(a_n\mp b_{n,m}e^{-i\omega (n-m)t})|
\end{eqnarray}
with coefficients $a_n=\langle n_A|\alpha \rangle ^2+\langle
n_B|\alpha \rangle ^2,b_{n,m}= \langle n_A|\alpha \rangle  \langle
n_B|\alpha \rangle (\langle m_B|n\rangle_A +\langle m_A|n\rangle_B
)$. Therefore, the atomic population inversion for strong coupling
strength is approximately obtained as
\begin{equation}\label{analytical}
P(t)\simeq
\sum_{nm}\frac{-2a_{n}b_{n,m}}{\sqrt{a_{n}^2+b_{n,m}^2}}\cos
[\omega (n-m)t]
\end{equation}
It is just the summation of $\texttt{cos}[\omega(n-m)t]$ that
results in the recovery of the collapses and revivals, similar to
the case of the RWA described by  Eq.(~\ref{PRWA}). By using
Eq.(~\ref{analytical}), the strong coupling perturbative results
with the red curve is also exhibited in Fig.~\ref{comp}(d),  which
almost superpose with the black curves. This reappearance of the
collapses and revivals is originated from the nearly equal level
spacing  for the dressed eigenstates in the strong coupling regime,
as also demonstrated in Fig.~\ref{energy}.

In summary, within the bosonic coherent-state approach, two
interesting features of the JC system well studied previously are
examined in this paper.  Even in the present experimentally
accessible regime, say, $g/\omega=0.1$, the height of the two peaks
of the vacuum Rabi splitting are different, the collapses and
revivals in the evolution of the population inversion completely
disappear. In the deep strong coupling regime, e.g. g=2, the  vacuum
Rabi splitting vanish and the collapses and revivals in the
evolution of the atomic population inversion reappear. Further
experiments are then motivated to confirm these theoretical
predictions.

This work is supported by RGC and FRG. The work of CQH was supported
in part by National Basic Research Program of China under Grants No.
No. 2009CB929104. The work of ZYY was supported in part by the
Fundamental Research Funds for the Central Universities under Grant
No. CDJRC10300006.

\end{document}